# Les produits Halal dans les économies occidentales

**Abdelatif KERZABI**

Professeur

Faculté des sciences économiques et de gestion

Université de Tlemcen  ALGERIE

Mail : kerzabi57@gmail.com

**Résumé :**

Depuis quelques années, on entend parler des produits halals et cela dans les sociétés non musulmanes notamment européennes et américaines. En France, pour ne pas citer d'autres exemples, le chiffre d'affaires des produits halals vendus en magasin en l'an 2010, a augmenté de 23% et représente 5,5 milliards d'euros dont 1,1 milliards pour la restauration rapide[1], et ça n'a pas cessé d'augmenter depuis. Un nouveau marché qui ne concerne pas seulement les musulmans. Le halal a une interprétation religieuse mais son développement rapide exige que l'on s'interroge sur ses véritables motivations qui ne sont pas seulement religieuses

**Abstract :**

In last years, we hear about halal products in non- Muslim societies including European and American ones. In France, for example, sales of halal products sold in stores during the year 2010, increased 23 % and represented 5.5 billion euros, including 1.1 billion for the fast food, and it has not stopped growing since. A new market is not only about Muslims. Halal is a religious interpretation but its rapid development requires that we question his true motives are not just religious

Mots clés : Halal, marketing, religion, occident, islam.

Dans ce présent travail, nous tacherons de répondre aux questions suivantes :

**Qu'est ce que signifie précisément le terme Halal? Comment est-il apparut,**

---

[1] D'après une étude du cabinet de conseil Insights SymphonyIRI Group.

**et faut-il considérer comme non démocratiques les comportements et institutions qui serait inspirés par des croyances religieuses ?**

1. **Religion en occident :**

Sous l'effet de groupes de pression ou d'influence religieux, plusieurs Etats démocratiques occidentaux revendiquent explicitement leur attachement à des traditions confessionnelles (Italie, Espagne, Pologne…). L'identité religieuse au sein de la France cosmopolite suscite des critiques envers la rationalité laïque des institutions politiques. De ce fait, le religieux n'a jamais succombé à la laïcité. Il a toujours été présent.

**1.1 Religion et postmodernité**

La modernité (Henni, 2006), est ce mouvement de renversement des sociétés d'ordres et de statuts. Idéalement, elle fait émerger le libre devenir individuel et supprime toute prédétermination de nature, de parenté ou de statut. Le principe de la modernité est : on devient homme en liquidant toute dette pour ne rien devoir à personne. Ceci suppose qu'il n'y ait aucune souveraineté supérieure ou appartenance de nature ou de statut qui puisse faire et défaire les destinées individuelles (Henni, 2006). Aujourd'hui, les éléments de cette modernité sont remis en cause par un retour à des sociétés de statuts et de rentes, localement et mondialement. Ces sociétés de statuts, vont favoriser l'émergence de communautés de types religieuses.

A partir du XVIIIè siecle, l'institutionnalisation du pouvoir et du droit a privatisé le rapport à la religion. A l'inverse de l'héritage français qui a conduit à un affrontement entre la morale publique et la morale privée, la filiation américaine, illustre une vision originale dans le sens où le retrait de la religion hors de la vie publique a renforcé le repli des citoyens sur des communautés religieuses, unifiées par des valeurs morales traditionnelles. Le partage de ces valeurs communes par ces communautés a permis la circulation d'un bien commun d'inspiration chrétienne. Ce facteur religieux a justement favorisé la démocratisation de la société américaine.

Pour un spécialiste des mouvements islamistes, J. Kepel (1991) fut le premier à se poser la question du retour du religieux. Cet auteur va s'intéresser à trois types de fondamentalisme : l'islamisme radical dans les pays musulmans ; le militantisme protestant avec l'évangélisme conservateur américain ; et le retour au judaïsme et à l'observance intégrale de la loi biblique qui s'affichait dans les communautés juives du monde entier.

Le renouveau de l'islam qui intéresse nos propos, intervient suite à l'échec des alternatives marxistes et nationalistes. L'islamisme prend le relai de ces idéologies nationalistes et marxistes. Armé de ses dogmes et de sa promesse millénariste, l'islamisme politique fait irruption pour proposer une alternative à la communauté des croyants qui transcende le dogme libéral imposé par l'impérialisme occidental. L'adhésion des populations à ce discours qui promet le salut sur terre ou au ciel leur apporte de réels bénéfices tant sur le plan psychologique que matériel. Ce discours se ressource dans les milieux populaires qui leur donne la possibilité de s'exprimer alors qu'ils se considéraient avant comme marginalisé et laissés pour compte.

La religion au-delà des croyances, devient une réponse aux angoisses (Onfray, 2005 ) et sert en même temps a affronter les difficultés de la vie. De là on peut considérer que le retour au religieux est une réponse à la postmodernité.

**1.2 Un phénomène d'identification communautaire**

En Occident, quand il s'agit de l'autre, le non autochtone, nous retenons de l'histoire que l'inégalité sociale se transforme en inégalité raciale (Henni, 2006). Samuel Huntington, auteur du « Clash des civilisations » proclame la supériorité de la société américaine par la culture protestante, la prééminence de la loi et le droit des individus. C'est ainsi que cet auteur, généralise son idée sur l'ensemble de l'occident pour marquer sa hauteur par rapport à l'Orient. Ici, il s'agit bien sur du monde musulman. Le patrimoine défendu par Huntington est civilisation. Les sociétés ou les individus étrangers à ce patrimoine sont considérés comme non civilisé. La non appartenance à ce patrimoine «civilisationnel » signifie un rejet de l'autre. Cependant, nous considérons que ce rejet de l'autre est propre aux mutations du capitalisme depuis la fin des années 1970. Les philosophes vont appeler ce phénomène de postmodernité.

Ce rejet de l'autre qui se fait au nom de la civilisation occidentale va produire un repli identitaire des communautés musulmanes vivant en Occident. Les individus, ont tendance à se communautariser pour faire face a l'inégalité des origines qui prend une forme raciale. C'est ainsi que le turc et le maghrébin ne peuvent prétendre aux mêmes statuts que ceux des européens d'origine.

Cette situation nouvelle est provoquée par les mutations du capitalisme. Aujourd'hui, face au retrait massif de l'Etat providence causé par la libéralisation croissante depuis les années 1980, face au chômage, à une pauvreté croissante, à une modernisation de plus en plus insatisfaisante incapable de créer des liens de solidarité, les gens cherchent de plus en plus le réconfort dans la religion. Les laissés pour compte, écoutent avec intérêt les marchands de salut qui leur promettent le paradis.

Au plan économique, la modernité est native de la révolution industrielle. Elle est fondé sur la croyance dans le progrès de la science et de la technique, et donc aux bienfaits de l'industrie pour le bien commun.

A partir des années 1980, au plan économique, la production de masse cède la place à une production plus individualisée. L'automatisation de la production supprime les emplois et abaisse les couts de production. De nouveaux produits apparaissent : ordinateurs, téléphones mobiles, baladeurs, DVD etc. Ces produits répondent tous à un besoin de communication et de distraction, voire à un désir d'évasion dans le monde de l'imaginaire.

   2. **Le produit halal : quelle définition ?**

Pour l'Islam, le produit halal est tout produit dont la consommation est permise par la loi islamique. Celle-ci est l'ensemble des préceptes de l'Islam, qu'ils émanent des textes coraniques, des paroles du prophète, ou du conseil des musulmans. Ainsi, sont exclus de la consommation halal, les boissons alcoolisées, le porc et la viande dont l'animal n'est pas égorgé selon le sacrifice rituel et par un non musulman.

Le halal est définit par rapport à la norme Islamique. Cette norme prévoit les interdits et ce qui est toléré. Sur le plan juridique, selon le CODEX ALIMENTARIUS, l'aliment halal est tout aliment autorisé par la Loi islamique, et qui répond aux conditions ci-après :

-il ne doit ni constituer ni contenir quoi que ce soit jugé illégal conformément à la Loi islamique;

- il ne doit pas avoir été préparé, transformé, transporté ou entreposé à l'aide d'instruments ou d'installations non conformes à la Loi islamique;

-au cours de sa préparation, de sa transformation, de son transport ou de son entreposage, il ne doit pas avoir été en contact direct avec des aliments ne répondant pas aux deux premières conditions ci-dessus.

Les aliments conformes à la loi islamique sont tout aliment de toute origine sauf ceux qui proviennent des animaux et plantes ci-après et des produits qui en dérivent :

## 2.1 Aliments d'origine animale

(a) porcs et sangliers,

(b) chiens, serpents et singes,

(c) animaux carnivores munis de griffes et de crocs comme le lion, le tigre, l'ours, etc.,

(d) oiseaux de proie munis de serres comme les aigles, les vautours, etc.,

(e) ravageurs tels que rats, mille-pattes, scorpions, etc.,

(f) animaux qu'il est interdit de tuer en Islam, par exemple fourmis, abeilles et piverts,

(g) animaux jugés généralement répugnants tels que poux, mouches, vers de terre, etc.,

(h) animaux qui vivent aussi bien sur terre que dans l'eau tels que grenouilles, crocodiles, etc.,

(i) mulets et ânes domestiques,

(j) tous les animaux aquatiques venimeux et dangereux,

(k) tout autre animal abattu selon des méthodes non conformes à la Loi islamique,

(l) sang.

**2.2 Aliments d'origine végétale et autres produits**

a) Plantes toxiques et dangereuses sauf quand la toxine ou le danger peuvent être éliminés durant la transformation.

b) boissons alcoolisées, enivrantes ou dangereuses;

c) Les additifs alimentaires obtenus à partir des produits cités plus haut

Enfin, l'abattage de tous les animaux dont la consommation est autorisée par la loi devrait être abattus conformément aux règles énoncées dans le Code d'usages Codex recommandé en matière d'hygiène pour les viandes fraîches. La personne chargée de l'abattage doit être un musulman sain d'esprit et connaissant bien les méthodes d'abattage de l'Islam;

### 3. Quantification de l'économie Halal

Le Halal ou licite en arabe représente un marché non négligeable pour les entreprises, notamment celles qui exportent. Les musulmans, au nombre de 1,6 Milliards de consommateurs potentiels dans le monde est un marché Halal de presque 450 milliards d'euros. Ce marché, en croissance de 10%, attire de plus en plus d'entreprises qui se tournent vers les pays musulmans. Les produits Halal deviennent de plus en plus nombreux, ils concernent non seulement les produits carnés où la certification halal est obligatoire mais aussi les produits cosmétique, les médicaments ainsi que les compléments alimentaires et les plats préparés.

### 3.1 Le halal, un secteur en expansion

Aujourd'hui, l'économie Halal à largement dépassé les produits de la boucherie pour s'étendre à une large gamme de produits. On y trouve, les parfums, les fromages et même les crédits bancaires. Le marché de l'alimentation selon le magasine « Time » est le plus concerné par le Halal puisqu'il représente un chiffre d'affaires de plus de 600 Milliards de dollars par an soit 16% de l'industrie agroalimentaire mondiale. Cette ruée vers le secteur de l'alimentation est motivée par la population musulmane qui connait des taux de croissance soutenus et à laquelle il faut répondre par des produits correspondant au rite musulman. C'est ainsi, que plusieurs grandes enseignes mondiales telles que Nestlé et McDonald's ont développé des produits pour ce marché et arrivent à contrôler 90% des produits de ce secteur. Aussi, la grande distribution tente d'attirer une clientèle qui, jusqu'à présent, achète prioritairement dans les commerces traditionnels. Cette clientèle est une population jeune moins de 40 ans et qui est habituée à la grande surface ainsi qu'à la restauration rapide. Selon le magazine « Le point[2] », 71 % des Français d'origine maghrébine fréquentaient les fast-foods, les sandwicheries et les kebabs. On achète plus souvent un hamburger qu'un pantalon.

**3.2 : A qui profite le halal ?**

Aujourd'hui, le halal s'est développé de façon exponentielle. Beaucoup d'acteurs économiques ont récupéré et instrumentalisé ce précepte religieux pour en faire un business. Retranché uniquement à la viande il y a quelques années, on assiste actuellement à un retour du halal qui témoigne d'un repli identitaire que l'on constate dans d'autres comportements, comme le montre l'extension du port du voile qui n'était qu'un phénomène sans importance.

L'intervention massive des acteurs économiques et notamment la grande distribution a certainement contribué à accentuer ce communautarisme.

Les exportations bretonnes représentent aujourd'hui 210 M€ (8,16% des exportations bretonnes) vers les pays musulmans, principalement l'Arabie saoudite, les Émirats arabes unis et la Malaisie.

---

[2] Le Point.fr - Publié le 17/02/2010

Des entreprises de transports (qui n'embarquent pas de porc ou d'alcool), mais aussi des chaînes d'hôtels (avec notamment des piscines qui écartent la mixité) se sont lancées dans les pays musulmans et lorgnent sur l'Europe. L'Angleterre devrait bientôt ouvrir son premier hôtel halal.

La toute-puissante industrie de la viande, dont les entreprises Bigard, Doux, Duc, LDC (Celvia), Panzani (Zakia), pour les plus connus, engrange d'importants bénéfices chaque année grâce à la filière halal.

Ce marché échappe aux mains des industriels et entrepreneurs musulmans. Parmi les dix plus grands pays producteurs de viande halal, aucun n'est un pays musulman, 90% des bénéfices du marché halal vont en Occident. Ces produits n'ont pu être vulgarisés que dans les pays occidentaux qui ont fini par prendre l'avantage sur les pays musulmans. A l'origine, des Musulmans, minoritaires dans ces pays occidentaux, avaient du mal à trouver des produits conformes aux principes islamiques.

Conclusion :

La modernité est un mode de civilisation caractéristique qui s'oppose au mode de la tradition, c'est à dire à toutes les autres cultures antérieures ou traditionnelles. De ce fait, la modernité doit s'opposer à la religion. Nous constatons aujourd'hui que ce n'est pas le cas. La religion s'adapte à la modernité. Le halal et son émergence dans les pays qui déclarent leur modernité en est une preuve. Nous pensons même que le Halal qui est une manifestation du religieux s'est très bien insérer dans le monde moderne. La consommation des produits halals en nette progression signifie un retour au communautarisme qui est un refuge à une mondialisation qui fragilise les Etats. Le recul de l'Etat sous l'effet de la mondialisation a produit une identification des individus par le religieux. Cette identification se traduit par un ensemble de comportements dont la consommation du Halal.

Enfin, nous pouvons avancer que la religion ne disparaîtra pas du monde moderne car les hommes ont besoin de garder leur identité, leur mémoire, leurs racines. Cependant, pour subsister et faire face à un monde en pleine évolution, la religion devra se constituer et se reformer pour pouvoir s'adapter et construire de nouvelles mémoires.

## Références